\documentclass[aps,showpacs]{revtex4}
\usepackage{graphicx}
\def\no{\noindent}
\def\bc{\begin{center}}
\def\ec{\end{center}}

\def\beq{\begin{equation}}
\def\eeq{\end{equation}}

\begin{document}


\title{Diffusion in the random gap model of mono- and bilayer graphene 
}

\author{K. Ziegler}
\affiliation{Institut f\"ur Physik, Universit\"at Augsburg\\
D-86135 Augsburg, Germany}
\date{\today}

\begin{abstract}
In this paper we study the effect of a fluctuating gap in mono- and bilayer graphene, 
created by a symmetry-breaking random potential. We identify a continuous symmetry
for the two-particle Green's function which is spontaneously broken in the
average two-particle Green's function and leads to a massless fermion mode. 
Within a loop expansion it is shown that the massless mode is dominated on large
scales by small loops. This result indicates diffusion of electrons. 
Although the diffusion mechanism is the same in mono- and in bilayer graphene, 
the amount of scattering is much stronger in the latter.
Physical quantities  at the neutrality point, such as the density of states, 
the diffusion coefficient and the conductivity, are determined by the one-particle
scattering rate. All these quantities vanish at a critical value of the average
symmetry-breaking potential, signaling a continuous transition to an insulating behavior.
\end{abstract}

\pacs{81.05.Uw,71.55.Ak,72.10.Bg,73.20.Jc}

\maketitle

\section{Introduction}

Graphene is a single sheet of carbon atoms, where the latter form a honeycomb lattice. 
Graphene as well as a stack of two graphene sheets (i.e. a graphene bilayer) are semimetals 
with remarkably good conducting properties 
\cite{novoselov05,zhang05,geim07}. These materials have been experimentally realized with external gates,
which allow a continuous change of charge carriers. There exists a non-zero minimal conductivity
at the charge neutrality point (NP). Its value is very robust and almost unaffected by disorder or 
thermal fluctuations \cite{geim07,tan07,chen08,morozov08}.

Many technological applications of graphene require an electronic gap to construct switching devices.
A first step in this direction has been achieved by recent experiments with hydrogenated graphene 
\cite{elias08} and gated 
bilayer graphene \cite{ohta06,oostinga08,gorbachev08}. These experiments take advantage of the fact 
that the breaking of a discrete symmetry of the lattice system opens a gap in the elctronic spectrum
at the Fermi energy. A symmetry-breaking potential (SBP)
is a staggered potential in the case of a monolayer, which breaks the sublattice symmetry of the
honeycomb lattice, or a gate potential that distinguishes between the two layers in the case of bilayer
graphene, where the latter breaks the symmetry between the layers.
With this opportunity, one enters a new field, where one can switch 
between conducting and insulating regimes of a two-dimensional material, 
either by a chemical process (e.g. oxidation or hydrogenation) or by applying an 
external electric field \cite{castro08}.

The opening of a {\it uniform} gap destroys the metallic state immediately. Thus the conductivity of the 
material would drop from a finite value of order $e^2/h$ directly to zero. In a realistic system, 
however, the gap may not be uniform after turning on the SBP. This means that only locally the material becomes
insulating, whereas in other regions of the sample it is still metallic. The situation can be compared with a 
classical random network of broken and unbroken bonds. 
The conductivity of such a network is nonzero as long as there is a percolating cluster of unbroken
bonds. In such a system the transition from conducting to insulting behavior is presumably
a second order percolation transition \cite{cheianov07}.

Disorder in mono- and bilayer graphene has been the subject of a number of recent numerical studies 
\cite{xiong07,zhang08} and analytic calculations \cite{dcconductivity,ziegler97,ziegler08}.
The results can be summarized by the statement that chiral-symmetry 
preserving disorder provides delocalized states whereas a chiral-symmetry breaking scalar potential 
disorder leads to Anderson localization, even at the NP. The  breaks the chiral symmetry
but still allows for delocalized states at the NP \cite{ziegler97,ziegler09a}. In contrast to
chiral-symmetry preserving disorder, a random SBP reduces the minimal conductivity and can even
lead to an insulating behavior.

In this article an approach will be employed that eliminates a part of the complexity of
the tight-binding model by focusing on continuous symmetries and corresponding spontaneous symmetry breaking. 
This allows us to identify a massless mode in the system with a randomly fluctuating SBP.
Using a loop expansion we study the scaling behavior of the model and derive a diffusion
propagator for the asymptotic behavior on large scales. Our result implies a relation between
the average two-particle Green's function and the product two average one-particle Green's functions
in self-consistent Born approximation. This is similar to the solution of the Bethe-Salpeter 
equation, a self-consistent equation for the average two-particle Green's function (Cooperon) 
\cite{suzuura02,peres06,khveshchenko06,mccann06,yan08}. In addition to the latter we are also
able to control the scaling behavior of all higher order terms in the loop expansion.

Our approach provides also information about the effect of symmetry-breaking terms. 
It turns out that the latter create a finite length scale $L_{\rm diff}$,
such that diffusion breaks down for length scales $L$ larger than $L_{\rm diff}$.
Another reason for the breakdown of diffusion is a vanishing spontaneous symmetry breaking.
This happens when the average value of the SBP exceeds a critical value. In this case
there is no drop of the conductivity but a continuous decay to zero, depending on 
the fluctuations of the SBP. 

This article is organized as follows. In Sect. \ref{sect2} the model and functional-integral
representation of the Green's functions are introduced. The symmetries of the model are
discussed in Sect. \ref{symmetry000}. Then an effective functional integral is constructed
for the average two-particle Green's function (Sect. \ref{averagetheory}) and a saddle-point
approximation is employed (Sect. \ref{saddlepoint}). The invariance of the saddle-point
equation of Sect. \ref{saddlepoint} under a continuous symmetry transformation requires 
the integration over a saddle-point manifold. This is discussed in detail in Sect. \ref{spmanifold},
which includes the loop expansion (Sect. \ref{loopexpansion}). The results of the loop
expansion and its consequences for the transport properties in graphene are discussed in Sect.
\ref{discussion}.  Finally, we conclude with a summary of our results in Sect. \ref{conclusions}.

\section{Model}
\label{sect2}

Quasiparticles in monolayer graphene (MLG) or in bilayer graphene (BLG) 
are described in tight-binding approximation by a nearest-neighbor hopping Hamiltonian 
\beq
{\bf H}=-{\sum_{<r,r'>}}t_{r,r'} c^\dagger_r c_{r'}
+\sum_r V_r c^\dagger_r c_r +h.c.
\ ,
\label{ham00}
\eeq
where $c_r^\dagger$ ($c_r$) are fermionic creation (annihilation) operators at lattice site $r$.
The underlying lattice structure is either a honeycomb lattice (MLG) or two honeycomb lattices 
with Bernal stacking (BLG) \cite{mccann06b,castro08}. There is an intralayer hopping rate $t$ and 
an interlayer hopping rate $t_\perp$ for BLG. $V_r$ is either a staggered potential (MLG) with $V_r=m$ on
sublattice A and $V_r=-m$ on sublattice B, or it is a biased gate
potential in BLG that is $V_r=m$ ($V_r=-m$) on the upper (lower) graphene sheet. These potentials
obviously break the sublattice symmetry of MLG and the symmetry between the two layers in BLG,
respectivly. A staggered potential can be the result of chemical absorption of other atoms in MLG
(e.g. oxygen or hydrogen \cite{elias08}). The potential in BLG has been realized as an 
external gate voltage, applied to the two layers of BLG \cite{ohta06}.
A consequence of the symmetry breaking is the formation of a gap $\Delta_g=m$ in both systems:
The spectrum of MLG consists of two bands with dispersion
\beq
E_k=\pm\sqrt{m^2+\epsilon_k^2} \ ,
\eeq
where
\beq
\epsilon_k^2=t^2[3+2\cos k_1+4\cos(k_1/2)\cos(\sqrt{3}k_2/2)]
\eeq
for lattice spacing $a=1$.
The spectrum of BLG consists of four bands \cite{castro08} with two low-energy bands 
\beq
E_k^-(m)=\pm\sqrt{\epsilon_k^2+t_\perp^2/2+m^2-\sqrt{t_\perp^4/4+(t_\perp^2+4m^2)\epsilon_k^2}}
\eeq
and two high-energy bands
\beq
E_k^+(m)=\pm\sqrt{\epsilon_k^2+t_\perp^2/2+m^2+\sqrt{t_\perp^4/4+(t_\perp^2+4m^2)\epsilon_k^2}} \ .
\eeq
The spectrum of the low-energy bands has nodes for $m=0$ where $E_k^-(0)$ vanishes. These nodes
are the same as those of a single layer. For small gating potential 
we can expand $E_k^-(m)$ under the square root near the nodes and get
\[
 E_k^-(m)\sim \pm\sqrt{m^2+E_k^-(0)^2}
 \]
with the same gap as in MLG.

The two bands in MLG and the two low-energy bands in BLG represent a spinor-1/2 wave
function. This allows us to expand the corresponding Hamiltonian in terms of
Pauli matrices $\sigma_j$ as
\beq
H=h_1\sigma_1+h_2\sigma_2+m\sigma_3 \ .
\label{ham01}
\eeq
Near each nodes the coefficients $h_j$ are \cite{mccann06}
\beq
h_j=i\nabla_j \ \ (MLG), \ \ h_1=\nabla_1^2-\nabla_2^2, \ h_2=2\nabla_1\nabla_2 \ \ (BLG) \ ,
\label{elements}
\eeq
where $(\nabla_1,\nabla_2)$ is the 2D gradient.

Neither in MLG nor in BLG the potential is uniform. The reason in the case of MLG is that
fluctuations appear in the coverage of the MLG by additional non-carbon atoms.
In the case of BLG it is crucial that the graphene sheets are not planar but create
ripples \cite{morozov06,meyer07,castroneto07b}. 
As a result, electrons experience a varying potential $V_r$ along each graphene
sheet, and $m$ in the Hamiltonian of Eq. (\ref{ham01}) is random variable in space.
For BLG it is assumed that the gate voltage is adjusted at 
the NP such that in average $m_r$ is exactly antisymmetric with respect to the two layers:
$\langle m_1\rangle_m=-\langle m_2\rangle_m$.

At first glance, the Hamiltonian
in Eq. (\ref{ham00}) is a standard hopping Hamiltonian with random potential $V_r$. This is a model
frequently used to study the generic case of Anderson localization \cite{anderson58}. The dispersion,
however, is special in the case of graphene due to the honeycomb lattice: at low energies it 
consists of two nodes (or valleys) $K$ and $K'$ \cite{castroneto07b,mccann06}. 
It is assumed here that weak disorder scatters only at small momentum such that 
intervalley scattering, which requires large momentum at least near the NP, is not relevant
and can be treated as a perturbation.
Then each valley contributes separately to transport, and the contribution of 
the two valleys to the conductivity $\sigma$ is additive:
$
\sigma=\sigma_K+\sigma_{K'}
$.
This allows us to consider the low-energy Hamiltonian in Eqs. (\ref{ham01}), (\ref{elements}),  
even in the presence of randomness for each valley separately. 
Within this approximation the term $m_r$ is a random variable with mean
value $\langle m_r\rangle_m ={\bar m}$ and variance 
$\langle (m_r-{\bar m})(m_{r'}-{\bar m})\rangle_m=g\delta_{r,r'}$.
The following transport calculations will be based entirely on the Hamiltonian of 
Eqs. (\ref{ham01}),(\ref{elements}). 
In particular, the average Hamiltonian $\langle H\rangle_m$
can be diagonalized by Fourier transformation and is
\[
\langle H\rangle_m =
k_1\sigma_1+k_2\sigma_2+{\bar m}\sigma_3
\]
 for MLG with eigenvalues $E_k=\pm\sqrt{{\bar m}^2+k^2}$.
For BGL the average Hamiltonian is
\[
\langle H\rangle_m =
(k_1^2-k_2^2)\sigma_1+2k_1k_2\sigma_2+{\bar m}\sigma_3
\]
with eigenvalues $E_k=\pm\sqrt{{\bar m}^2+k^4}$.

Transport properties of the model can be calculated from the Kubo formula.
Here we focus on interband scattering between states of energy $\omega/2$ and $-\omega/2$.
This is related to the zitterbewegung \cite{zitter}, 
which is a major contribution to transport near the NP. The frequency-dependent 
conductivity then reads \cite{ziegler08}
\beq
\sigma_0(\omega)
=-\frac{e^2}{2h}\omega^2 \langle\langle \Phi_{-\omega/2}|r_k^2|\Phi_{\omega/2}\rangle\rangle_m \ ,
\label{cond0b}
\eeq
where $|\Phi_{E}\rangle$ is the Fourier transform of the wave function under time 
evolution $\exp(-iHt)$:
\[
|\Phi_{E}\rangle\equiv\int_0^\infty e^{(iE-\epsilon)t}|\Psi(t)\rangle dt
=\int_0^\infty e^{(iE-\epsilon)t}e^{-iHt}dt|\Psi(0)\rangle
\]
\beq
=-i(H - E-i\epsilon)^{-1}|\Psi(0)\rangle =-iG(-E-i\epsilon)|\Psi(0)\rangle 
\label{evolution}
\eeq
with the one-particle Green's function $G(z)=(H+z)^{-1}$.
In other words, the conductivity is proportional to a matrix element of the position operator $r_k$
($k=1,2$) with respect to energy functions from the lower and the upper band. 
The matrix element 
on the right-hand side is identical with the two-particle Green's function:
\[
\langle \Phi_{-\omega/2}|r_k^2|\Phi_{\omega/2}\rangle
=\sum_r r_k^2 Tr_2\left[ G_{r0}(-\omega/2-i\epsilon)G_{0r}(\omega/2+i\epsilon)\right] \ .
\]
With the identity $H=-\sigma_n H^T\sigma_n$, where $n=1$ for MLG and $n=2$ for BLG
(cf discussion in Sect. \ref{symmetry000}), the matrix element also reads
\beq
\langle \Phi_{-\omega/2}|r_k^2|\Phi_{\omega/2}\rangle
=-\sum_r r_k^2 Tr_2\left[ \sigma_nG^T_{r0}(\omega/2+i\epsilon)\sigma_n G_{0r}(\omega/2+i\epsilon)\right] \ .
\label{cond2}
\eeq


\subsection{Functional Integral}

The two-particle Green's function on the right-hand side of Eq. (\ref{cond2})
can be expressed, before averaging, as a Gaussian functional integral with
two independent Gaussian fields, a boson (complex) field $\chi_{rk}$ and a fermion 
(Grassmann) field $\Psi_{rk}$ ($k=1,2$) and their conjugate counterparts ${\bar\chi}_{rk}$
and ${\bar\Psi}_{rk}$ \cite{negele}:
\beq
-G^T_{rr',jj'}(z)G_{r'r,k'k}(z)
=\int\Psi_{r'j'} {\bar \Psi}_{rj}\chi_{rk}{\bar \chi}_{r'k'} \exp(-S_0(z))
{\cal D}[\Psi] {\cal D}[\chi] \ .
\label{finta}
\eeq
$S_0(z)$ is a quadratic form of the four-component field 
$\phi_r=(\chi_{r1},\chi_{r1},\Psi_{r2},\Psi_{r2})$
\begin{equation}
S_0(z)=
-i\sum_{r,r'}
\phi_r\cdot({\hat H}+z)_{r,r'}{\bar\phi}_{r'} \ \ (Im z >0) \ ,
\label{ssa0}
\end{equation}
where the extended Hamiltonian ${\hat H}=diag(H,H^T)$ of $S_0$ acts 
in the boson and in the fermion sector separately.
The use of the mixed field $\phi_r$ has the advantage that an extra normalization
factor for the integral is avoided. 
The matrix element in Eq. (\ref{cond2}) reads now
\[
\langle \Phi_{-\omega/2}|r_k^2|\Phi_{\omega/2}\rangle
=\sum_{j\ne k}\sum_r r_k^2 \left[
\langle\Psi_{0j}{\bar\Psi}_{rj}\chi_{rk}{\bar\chi}_{0k}\rangle_0
-(-1)^n\langle\Psi_{0j}{\bar\Psi}_{rk}\chi_{rj}{\bar\chi}_{0k}\rangle_0
\right]
\]
\beq
=-\sum_{j\ne k}\sum_r r_k^2 
\left[
\langle\chi_{rk}{\bar\Psi}_{rj}\Psi_{0j}{\bar\chi}_{0k}\rangle_0
-(-1)^n\langle\chi_{rj}{\bar\Psi}_{rk}\Psi_{0j}{\bar\chi}_{0k}\rangle_0
\right]
\label{me2}
\eeq
with
\[
\langle ...\rangle_0
=\int ... \exp(-S_0(z)) {\cal D}[\Psi] {\cal D}[\chi] \ .
\]

\section{Symmetries}
\label{symmetry000}

Transport properties 
are controlled by the symmetry of the Hamiltonian and of the corresponding 
one-particle Green's function $G(i\epsilon)=(H+i\epsilon)^{-1}$. In the absence of
sublattice-symmetry breaking (i.e. for $m=0$), the Hamiltonian 
$H=h_1\sigma_1+h_2\sigma_2$ has a continuous chiral symmetry
\beq
H \to e^{\alpha\sigma_3} He^{\alpha\sigma_3}=H
\label{contsymmetry}
\eeq
with a continuous parameter $\alpha$, since $H$ anticommutes with $\sigma_3$.
The term $m\sigma_3$ breaks the continuous chiral symmetry. 
However, the behavior under transposition $h_j^T=-h_j$ for MLG and $h_j^T=h_j$ for 
BLG provides a discrete symmetry:
\beq
H\to -\sigma_n H^T\sigma_n =H \ ,
\label{discretesymm}
\eeq
where $n=1$ for MLG and $n=2$ for BLG.
This symmetry is broken for the one-particle Green's function $G(i\epsilon)$
by the $i\epsilon$ term. To see whether or not the symmetry is restored for $\epsilon\to0$,
the difference of $G(i\epsilon)$ and the transformed Green's function $-\sigma_nG^T(i\epsilon)\sigma_n$
must be evaluated:
\beq
G(i\epsilon)+\sigma_nG^T(i\epsilon)\sigma_n=G(i\epsilon)-G(-i\epsilon)  \ .
\label{op}
\eeq
For the diagonal elements this is the density of states at the NP $\rho(E=0)\equiv\rho_0$ 
in the limit $\epsilon\to0$ .
Thus the order parameter for spontaneous symmetry breaking is $\rho_0$.

Eq. (\ref{cond2}) indicates that transport properties are expressed by the two-particle 
Green's function
$G(i\epsilon)G(-i\epsilon)$. Each of the two Green's functions, $G(i\epsilon)$
and $G(-i\epsilon)$, can be considered as a random variable which
are correlated due to the common random variable $m_r$. Their distribution is
defined by a joint distribution function $P[G(i\epsilon),G(-i\epsilon)]$. 
In terms of transport theory, both Green's functions must be included
on equal footing. This is possible by introducing the extended Green's function
\beq
{\hat G}(i\epsilon)=\pmatrix{
G(i\epsilon) & 0 \cr
0 & G(-i\epsilon) \cr
} =\pmatrix{
H+i\epsilon & 0 \cr
0 & H-i\epsilon \cr
}^{-1} \ .
\label{extendedgf}
\eeq
In the present case one can use
the symmetry transformation of $H$ in Eq. (\ref{discretesymm}) to write 
the extended Green's function as
\[
{\hat G}(i\epsilon)=\pmatrix{
\sigma_0 & 0 \cr
0 & -\sigma_0 \cr
}
\pmatrix{
\sigma_0 & 0 \cr
0 & i\sigma_n \cr
}\pmatrix{
H+i\epsilon & 0 \cr
0 & H^T+i\epsilon \cr
}^{-1} \pmatrix{
\sigma_0 & 0 \cr
0 & i\sigma_n \cr
} \ .
\]
The extended Hamiltonian ${\hat H}=diag(H,H^T)$ is invariant 
under a global ``rotation'' 
\beq
{\hat H}\to e^{{\hat S}}{\hat H}e^{{\hat S}}={\hat H} \ , \ \ \ \ {\hat S}=\pmatrix{
0 & \alpha \sigma_n \cr
\alpha'\sigma_n & 0 \cr
}
\label{symmetry2}
\eeq
with continuous parameters $\alpha,\alpha'$. The invariance is a consequence of the fact that 
${\hat H}$ anticommutes with $S$. The $i\epsilon$ term of the Green's function also breaks this symmetry.
For $\alpha\alpha'=-\pi^2/4$ the diagonal element of ${\hat G}-e^S{\hat G} e^S$ is proportional
to the density of states $\rho_0$. Thus, the continuous
symmetry is spontaneously broken for  $\epsilon\to0$ if $\rho_0$ is nonzero. In this case there is 
a massless mode.

As a symmetry-breaking parameter, $\epsilon$ generates a characteristic response of the system with 
long-range correlations when it is varied for $\epsilon\sim0$ . This is reministent of a weak 
external magnetic field in a (classical) ferromagnet, where the response to a change of the 
magnetic field creates a power-law magnetic susceptibility near the critical point. Moreover, 
if $\epsilon$ is chosen
as a space-dependent field $\epsilon_r$, we can vary it locally and obtain a space-dependent
response in form of correlation functions of the Green's functions. This allows us to study
complex correlation functions by taking local derivatives of the field $\epsilon_r$. 

Returning to the quadratic form in the action $S_0(z)$ of Eq. (\ref{ssa0}), we notice that after the
``rotation'' of the $diag(H,H^T)$ with $e^{{\hat S}}$
off-diagonal block matrices are generated. These matrices should have Grassmann elements
in order to have a quadratic form that has pairs of complex and pairs of Grassmann variables. 
Therefore, the parameters $\alpha$ and $\alpha'$ must be Grassmann variables:
$\alpha=\psi$ and  $\alpha'={\bar\psi}$.

\section{Averaged matrix elements}
\label{averagetheory}

As an example, we need to consider the averaged matrix element of $r_k^2$ in Eq. (\ref{me2}).
Averaging Eq. (\ref{finta}) over the Gaussian distribution of $v_r$ means
replacing $\exp(-S_0)$ by $\langle \exp(-S_0)\rangle_m$ on the right-hand side of 
the equation. The latter can be written again as an exponential function
$\langle \exp(-S_0)\rangle_m=\exp(-S_1)$, where the new function $S_1$ contains also
quartic terms of the field $\phi$:
\beq
S_1=-i\sum_{r,r'}\phi_r\cdot(H_0+z)_{r,r'}{\bar\phi}_{r'}
+g\sum_r (\phi_r\cdot\gamma_3{\bar\phi}_r)^2 \ .
\label{effaction}
\eeq
Then it is convenient to transform the integration variables 
(Hubbard-Stratonovich transformation \cite{negele}) as 
\beq
\pmatrix{\chi_r{\bar\chi}_r&\chi_r{\bar\Psi}_r\cr
\Psi_r{\bar\chi}_r&\Psi_r{\bar\Psi}_r\cr
}\rightarrow
{\hat Q}_r=\pmatrix{
Q_r & \Theta_r\cr
{\bar\Theta}_r & -iP_r\cr
} \ ,
\eeq
where $Q_{r}$, $P_{r}$ are symmetric $2\times2$ matrices 
and $\Theta_{r}$, ${\bar\Theta}_{r}$
are $2\times2$ matrices whose elements are independent Grassmann variables. 
Now the correlation functions in Eq. (\ref{me2}) can be rewritten as correlation functions in the new field ${\hat Q}_r$.
Then the matrix element reads
\beq
\langle\langle \Phi_{-\omega/2}|r_k^2|\Phi_{\omega/2}\rangle\rangle_m
=-\frac{1}{g^2}\sum_{j\ne k}\sum_r r_k^2 \left[
\langle(\Theta\sigma_3)_{r,jk}({\bar\Theta}\sigma_3)_{0,kj}\rangle_2
-(-1)^n\langle(\Theta\sigma_3)_{r,jk}({\bar\Theta}\sigma_3)_{0,jk}\rangle_2
\right]
\label{me3}
\eeq
with
\[
\langle ...\rangle_2
=\int ... \exp(-S_2(z)) {\cal D}\Psi {\cal D}[{\hat Q}]
\]
and
\beq
S_2(z)=\sum_{r,r'}\frac{1}{g}{\rm Trg}({\hat Q}_{r}^2) 
+\ln [{\rm detg}[\langle{\hat H}\rangle_m+z -2\gamma_3{\hat Q}]] \ .
\label{action2}
\eeq
${\rm Trg}$ is the graded trace
\[
{\rm Trg}\left( \pmatrix{
A & \Theta \cr
{\bar \Theta} & B \cr
}\right)=Tr A - Tr B \ ,
\]
${\rm Tr}$ is the conventional trace, and ${\rm detg}$ is the graded determinant \cite{ziegler97}:
\beq
{\rm detg}\left( \pmatrix{
A & \Theta \cr
{\bar \Theta} & B \cr
}\right)=\frac{det(A)}{det(B)}det({\bf 1}-{\bar \Theta}B^{-1}\Theta A^{-1}) 
=\frac{det(A-{\bar \Theta}B^{-1}\Theta)}{det(B)}
\ .
\label{detg}
\eeq

\subsection{Saddle-point approximation}
\label{saddlepoint}

The integration in Eq. (\ref{me3}) can be
performed in saddle-point approximation. The saddle point is obtained as the
solution of $\delta S_2=0$. Assuming a solution of the form
\beq
{\hat Q}_0=-i\frac{\eta}{2}\gamma_3-\frac{m_s}{2}\gamma_0 \ ,
\label{SP00}
\eeq
we obtain the parameters $\eta$, $m_s$ from the saddle-point equation
\beq
{\hat Q}_0=g(\langle{\hat H}\rangle_m+z-2\gamma_3{\hat Q}_0)^{-1}_{rr}\gamma_3 \ .
\label{SPE00}
\eeq
A consequence of the symmetry discussed in Sect. \ref{symmetry000} is that
for $z=0$ the saddle-point equation is invariant under the global symmetry 
transformation ${\hat Q}_0\to{\hat U}^{-1}{\hat Q}_0{\hat U}$, where ${\hat U}=e^{\hat S}$ of Eq. (\ref{symmetry2}).
This transformation creates the saddle-point manifold
\beq
{\hat Q}_r'=-i\frac{\eta}{2}\gamma_3{\hat U}_r^2 -\frac{m_s}{2}\gamma_0 \ ,
\label{spm}
\eeq
where ${\hat U}_r$ is obtained from Eq. (\ref{symmetry2})
by replacing the transformation parameters $\alpha$ ($\alpha'$) by space-dependent Grassmann
variables  $\psi_r$ (${\bar\psi}_r$), respectively.
The form of ${\hat Q}_r'$, which is dictated by the symmetry, 
implies for the action $S_2$ on the saddle-point manifold that (i) the quadratic term
vanishes and (ii) the remaining term becomes
\beq
S'=\ln {\rm detg}(\langle{\hat H}\rangle_m+m_s\sigma_3+z+i\eta{\hat U}^2)
\ .
\label{spaction}
\eeq
This action contains the symmetry breaking field $z$. The matrix element of Eq. (\ref{me3}) becomes
\beq
\langle\langle \Phi_{-\omega/2}|r_k^2|\Phi_{\omega/2}\rangle\rangle_m
\approx \frac{4\eta^2}{g^2}\sum_r r_k^2 
\langle\psi_r{\bar\psi}_0\rangle_{S'}
\label{me4}
\eeq
with
\beq
\langle ... \rangle_{S'}=\int ... e^{-S'}{\cal D}[{\hat U}] 
=\int ... e^{-S'}{\cal D}[\psi] \ .
\label{average3}
\eeq
There is no extra factor from the invariant integration measure when we replace ${\cal D}[{\hat U}] $ by 
${\cal D}[\psi]$ (cf Appendix \ref{invmeas}).

\subsection{Evaluation of the scattering rate $\eta$}

The saddle-point approximation of the average one-particle Green's function means
\beq
\langle G(z)\rangle_m=\langle (H+z)^{-1}\rangle_m \approx (\langle H\rangle_m+m_s\sigma_3 +z+i\eta)^{-1} 
=G_{0}(z+i\eta) \ ,
\label{scba00}
\eeq
which is often called self-consistent Born approximation \cite{suzuura02}.
The ansatz for a uniform saddle-point solution in Eq. (\ref{SP00}) leads to
a shift of $z$ as $z\to i\eta'\equiv i\eta+z$ with
\begin{equation}
\eta'+iz =\eta' g I
\label{spea}
\end{equation}
and a shift of the average mass ${\bar m}\to {\bar m}+m_s$ with
\begin{equation}
m_s=-{\bar m}gI/(1+gI) \ .
\label{speb}
\end{equation}
The integral $I$ reads
\[
I=2\int G_{0,11}(i\eta')d^2k/(2\pi)^2/(i\eta')
\]
which is in the case of MLG
\begin{equation}
I\sim \frac{1}{\pi}\int_0^\lambda({\eta'}^2+({\bar m}+m_s)^2+k^2)^{-1}kdk
=\frac{1}{2\pi}\ln\left[ 1+\frac{\lambda^2}{{\eta'}^2 +({\bar m}+m_s)^2}\right]
\label{int1}
\end{equation}
and in the case of BLG
\begin{equation}
I\sim {1 \over \pi}\int_0^\lambda({\eta'}^2+({\bar m}+m_s)^2+k^4)^{-1}kdk
=\frac{\arctan\left(\lambda^2/\sqrt{{\eta'}^2+({\bar m}+m_s)^2}\right)}
{2\pi\sqrt{{\eta'}^2+({\bar m}+m_s)^2}} 
\sim \frac{1}{4\sqrt{{\eta'}^2+({\bar m}+m_s)^2}}
\label{int2}
\end{equation}
for $\lambda\sim\infty$.

A nonzero solution $\eta$ for $z=0$ requires $gI=1$ in Eq. (\ref{spea}), 
such that $m_s=-{\bar m}/2$ from Eq. (\ref{speb}). 
Since the integrals $I$ are monotonically decreasing functions for large ${\bar m}$,
a real solution with $gI=1$ exists only for $|{\bar m}|\le m_c$. For both physical
systems, MLG and BLG, the solutions read
\beq
\eta^2=(m_c^2-{\bar m}^2)\Theta(m_c^2-{\bar m}^2)/4 \ ,
\label{scattrate}
\eeq
where the model dependence enters only through the critical average parameter $m_c$:
\beq
m_c=\cases{
\frac{2\lambda}{\sqrt{e^{2\pi/g}-1}}\sim 2\lambda e^{-\pi/g} & (MLG) \cr
g/2 & (BLG) \cr
} \ .
\label{gap11}
\eeq
$m_c$ is much bigger for BGL, a result indicates that the effect of disorder 
is much stronger in BLG. This is also reflected by the scattering rate at ${\bar m}=0$ which 
is $\eta=m_c/2$.

\section{Integration over the saddle-point manifold}
\label{spmanifold}

The integration weight $\exp(-S')$ of the functional integral in Eq. (\ref{average3}) reads according 
to Eq. (\ref{spaction})
\beq
\exp(-S')={\rm detg}\left(H_0+i\epsilon +i\eta {\hat U}^2\right)^{-1}
\label{BW00}
\eeq
with the nonlinear field
\[
{\hat U}^2=e^{2{\hat S}}={\bf 1}+2{\hat S}+2{\hat S}^2
\]
and $H_0=\langle{\hat H}\rangle_m+m_s\sigma_3$.
We notice that
\[
{\bf 1}+{\hat S}+{\hat S}^2=({\bf 1}-{\hat S})^{-1} \ ,
\]
since ${\hat S}^l=0$ for $l\ge3$. This enables us to rewrite the integration weight as
\[
\exp(S')={\rm detg}\left(H_0+i\epsilon-i\eta +2i\eta ({\bf 1}-{\hat S})^{-1}\right)
={\rm detg}\left(({\bf 1}-{\hat S})(H_0+i\epsilon-i\eta) +2i\eta\right)
{\rm detg}({\bf 1}-{\hat S})^{-1}
\]
\beq
={\rm detg}\left({\bf 1}-{\hat S}(H_0+i\epsilon-i\eta)(H_0
+i\epsilon+i\eta)^{-1}\right){\rm detg}({\bf 1}-{\hat S})^{-1}\ ,
\label{weight}
\eeq
where we have used that ${\rm detg}(H_0+i\epsilon+i\eta)=1$.
This result 
is remarkable because (i) ${\hat S}$ appears only linearly in the determinants
and (ii) the matrix in the second determinant is diagonal:
\beq
{\rm detg}({\bf 1}-{\hat S})=\prod_r(1-2{\bar \psi}_r\psi_r) \ .
\label{measure}
\eeq
With the expression
\[
\delta{\hat G}_0:=(H_0+i\epsilon-i\eta)(H_0+i\epsilon+i\eta)^{-1}
={\bf 1}-2i\eta (H_0+i\epsilon+i\eta)^{-1}
\equiv{\bf 1}-2i\eta {\hat G}_0(i(\epsilon+\eta))\ ,
\]
we can write, using the definition of the graded determinant in Eq. (\ref{detg}), 
\[
\exp(-S')={\rm detg}\left({\bf 1}-{\hat S}\delta{\hat G}_0)\right)^{-1}\prod_r(1-2{\bar \psi}_r\psi_r) 
=det\left({\bf 1}-{\bar\psi}\sigma_1\delta G_{0,11}\sigma_1\psi\delta G_{0,22}\right)^{-1}
\prod_r(1-2{\bar \psi}_r\psi_r) \ .
\]

\no
$\delta{\hat G}_0$ depends on $\epsilon$, $\eta$ and satisfies for $n=1$ (MLG) or $n=2$ (BLG)
\[
\sigma_n\delta {\hat G}_{0,11}(\epsilon,\eta)\sigma_n=\sigma_n({\bf 1}-2i\eta G_{0,11}(i\epsilon+i\eta))\sigma_n
={\bf 1}+2i\eta G_{0,22}(-i\epsilon-i\eta)=\delta {\hat G}_{0,22}(-\epsilon,-\eta) \ .
\]
This implies for the integration weight
\beq
\exp(-S') 
=det\left({\bf 1}-{\bar\psi}h_-\psi h_+\right)^{-1}\prod_r(1-2{\bar \psi}_r\psi_r)
\label{weight3}
\eeq
with $h_\pm=\delta G_{0,22}(\pm\epsilon,\pm\eta)$, whose Fourier components are
\[
h_\pm\equiv \sigma_0\mp 2i\eta G_{0,22}(\pm i\epsilon\pm i\eta)
=\sigma_0\pm 2i\eta\sigma_n G_{0,11}(\mp i\epsilon\mp i\eta))\sigma_n
\]
\[
=\sigma_0\mp\frac{2i\eta}{(\eta+\epsilon)^2+h_1^2+h_2^2}\left(
\mp i(\eta+\epsilon)\sigma_0+(-1)^n(h_1\sigma_1-h_2\sigma_2)\right)
\]
\beq
=\left[1-\frac{2\eta(\epsilon+\eta)}{(\eta+\epsilon)^2+h_1^2+h_2^2}\right]\sigma_0
\pm\frac{2i\eta(-1)^n}{(\eta+\epsilon)^2+h_1^2+h_2^2}\left(-h_1\sigma_1+h_2\sigma_2\right) \ .
\label{hfourier}
\eeq

Eq. (\ref{weight3}) is probably the most compact representation of $\exp(-S')$, and a corresponding
simple visualization is that the lattice has isolated sites (due to
${\bar\psi}\psi\sigma_0$) or closed random walks of $h_+$ and $h_-$ pairs (due to ${\bar\psi}h_-\psi h_+$).
The functional integration in Eq. (\ref{average3}) can now be performed by expanding the determinant
$det\left({\bf 1}-{\bar\psi}h_-\psi h_+\right)^{-1}$ of Eq. (\ref{weight3}) in powers of the Grassmann
variables $\psi_r$ and ${\bar\psi}_r$. A nonzero contribution to the integral requires that the entire
lattice is covered with products $\psi_r{\bar\psi}_r$. This is quite different from the corresponding
functional integral with respect to complex fields, where already a single term of the expansion gives
a nonzero contribution. Consequently, the expansion must be organized in a specific way to control
the integration over the Grassmann variables. This can be done in terms of a loop expansion of the 
action $S'$ which is discussed in the next section.

\subsection{Loop expansion}
\label{loopexpansion}

\begin{figure}
\begin{center}
\includegraphics[width=7cm,height=5cm]{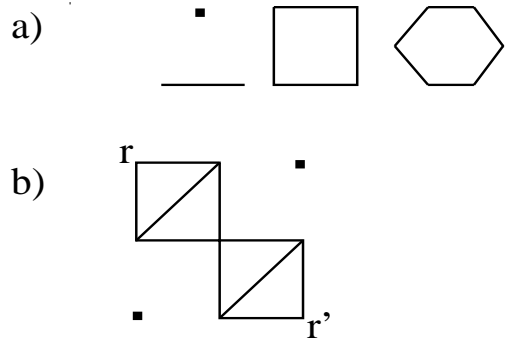}
\caption{
a) Elements of the loop expansion for the action $S'$ and b) for the two-particle Green's 
function $\langle G_{rr'}(z)G_{r'r}(-z)\rangle_m$. 
The dot corresponds with a simple factor $\psi_r{\bar\psi}_r$ from Eq. (\ref{measure}),
whereas the loops with $l$ corners correspond with the expansion term of order $l$
in Eq. (\ref{loopexp00}). Only an even number of corners can appear in the loop expansion a)
and each site must be visited twice by line elements in b), except for the end points $r$ and $r'$,
which are visited once.
}
\label{figloopexp}
\end{center}
\end{figure}
Starting from the expression in Eq. (\ref{weight3})
\[
det\left({\bf 1}-{\bar\psi}h_-\psi h_+\right)^{-1}
=\exp(-\ln det\left({\bf 1}-{\bar\psi}h_-\psi h_+\right))
\]
we can expand the exponent with trace terms of growing size as
\beq
\ln det\left({\bf 1}-{\bar\psi}h_-\psi h_+\right)
=-\sum_{l\ge 1}\frac{1}{l} Tr\left[({\bar\psi}h_-\psi h_+)^l\right]
=\sum_{l\ge 1}\frac{1}{l} Tr\left[(h_-\psi h_+{\bar\psi})^l\right] \ .
\label{loopexp00}
\eeq
The trace terms can be visualized as closed polygons (loops) on the lattice with alternating $\psi$ 
and ${\bar\psi}$ at the corners (cf Fig. \ref{figloopexp}a), 
where each term is normalized by the number of corners of the loop $l$. 
Inserting this in the functional integral of Eq. (\ref{average3}), all the loops can
contribute with the condition that they cover partially the lattice with products $\psi_r{\bar\psi}_r$.
There are many graphically equivalent coverages (but with different values), as can be seen in 
Fig. \ref{figloopexp}b: a square can either be a product of four $l=2$ contributions or just
one $l=4$ contribution. This equivalence raises the question for the
contribution(s) to a given graph with highest weight in the functional integral.  
A way to study this is a scaling analysis, where we analyse the change of the loop-expansion terms
under a change of length scales. For this purpose it
is convenient to choose the Fourier representation
\[
Tr\left[(h_-\psi h_+{\bar\psi})^l\right]
=\int ... \int Tr_2\left[
h_{-,k_1}\psi_{k_1-k_2}h_{+,k_2}{\bar\psi}_{k_2-k_3}\cdots
h_{-,k_{2l-1}}\psi_{k_{2l-1}-k_{2l}}h_{+,k_{2l}}{\bar\psi}_{k_{2l}-k_1}
\right]d^2k_1...d^2k_{2l} \ .
\]
It should be noticed that there are only $2l-1$ integrations that affect the field
$\psi$ and its conjugate, namely $k_1-k_2, k_2-k_3, ... ,k_{2l}-k_1$, since the sum of these
variables gives zero. The integration over the remaining $2l^{th}$ variable affects only the
$h$'s.
\no
Using $\Delta_j=k_j-k_{j+1}$ with $k_{2l+1}=k_1$ we get
\beq
Tr\left[(h_-\psi h_+{\bar\psi})^l\right]
=\int C_{\Delta_1,...,\Delta_{2l}}\psi_{\Delta_1}{\bar\psi}_{\Delta_2}
\cdots \psi_{\Delta_{2l-1}}{\bar\psi}_{\Delta_{2l}}\delta(\Delta_1+\dots +\Delta_{2l})
d^2\Delta_1\cdots d^2\Delta_{2l}
\label{loop1}
\eeq
with the coefficient
\beq
C_{\Delta_1,...,\Delta_{2l}}=\int Tr_2(h_{-,\Delta_1+\cdots +\Delta_{2l}+k_1}
h_{+,\Delta_2+\cdots +\Delta_{2l}+k_1}\cdots h_{-,\Delta_{2l-1}+\Delta_{2l}+k_1}h_{+,\Delta_{2l}+k_1}) d^2k_1
\ .
\label{coeffloop}
\eeq
These integral expressions contribute with different weight to the loop expansion of $\exp(-S')$, 
depending on the number of corners $l$. In order to analyse the weights we can use the fact that
$\Delta_j$ as well as $\psi_{\Delta_j}$ are integration variables in the functional integral.
This enables us to rescale them as
\beq
\Delta_j\to s^{-1} \Delta_j ,\ \ \psi_{\Delta_j}\to s^{-\alpha} \psi_{s\Delta_j}
\label{scaling00}
\eeq
and use the integration symbols as before the rescaling.
Then the scaling behavior of the general loop-expansion term in Eq. (\ref{loop1}) is
\[
\int C_{\Delta_1,...,\Delta_{2l}}\psi_{\Delta_1}{\bar\psi}_{\Delta_2}
\cdots \psi_{\Delta_{2l-1}}{\bar\psi}_{\Delta_{2l}}\delta(\Delta_1+\dots +\Delta_{2l})d^2\Delta_1\cdots 
d^2\Delta_{2l}
\]
\beq
\to
s^{2l(2+\alpha)}s^{-2}
\int C_{s\Delta_1,...,s\Delta_{2l}}\psi_{\Delta_1}{\bar\psi}_{\Delta_2}
\cdots \psi_{\Delta_{2l-1}}{\bar\psi}_{\Delta_{2l}}\delta(\Delta_1+\dots +\Delta_{2l})d^2\Delta_1\cdots 
d^2\Delta_{2l} \ .
\label{scaling2}
\eeq
Next, the contribution of $C_{s\Delta_1,...,s\Delta_{2l}}$ to the prefactor must be determined.
For $l=1$ we have $\Delta_2=-\Delta_1$ such that
\[
\int Tr_2(h_{-,k_1}\psi_{k_1-k_2}h_{+,k_2}{\bar\psi}_{k_2-k_1})d^2k_1 d^2 k_2
=\int \psi_{\Delta_1}\int Tr_2(h_{-,k_1}h_{+,-\Delta_1+k_1})d^2k_1{\bar\psi}_{-\Delta_1} d^2 \Delta_1
\]
\beq
\equiv \int \psi_{\Delta_1}C_{\Delta_1}{\bar\psi}_{-\Delta_1} d^2 \Delta_1 \ .
\label{1loop}
\eeq
This expression rescales as
\[
\int \psi_{k}C_{k}{\bar\psi}_{-k} d^2 k 
\to s^{2+2\alpha}\int \psi_{k}C_{sk}{\bar\psi}_{-k} d^2 k \ ,
\]
where $C_{sk}\approx C_0+s^2k^2C_0''$. Now we choose $\alpha=-2$ which gives a prefactor 1
for the $s^2k^2C_0''$ term. 

In general,
for $s<1$ the rescaling of the wavevector in Eq. (\ref{scaling00}) has the effect that 
the integration is shifted to larger
values in $\Delta_j$ (i.e. to shorter scales in real space). This is compensated by a prefactor in front of 
the integral. A prefactor smaller than 1 means that the integral contributes more on larger 
values of $k_j$ than on smaller values. In other words, the corresponding loop contributes more to 
shorter length scales than to larger ones. Since we are interested in large-scale properties, terms with
prefactors smaller than 1 are asymptotically irrelevant for this regime. 
The scaling of the coefficient 
\[
C_{s\Delta_1,...,s\Delta_{2l}}\sim \int Tr_2(h_{-,s\Delta_1+\cdots +s\Delta_{2l}+k_1}
h_{+,s\Delta_2+\cdots +s\Delta_{2l}+k_1}\cdots h_{-,s\Delta_{2l-1}+s\Delta_{2l}+k_1}h_{+,s\Delta_{2l}+k_1}) d^2k_1 \ ,
\]
for $l\ge 2$ can be studied by rescaling $h_\pm$.
Then we have for each factor $h_{\pm,s\Delta_j +\cdots +\Delta_{2l}+k_1}$
\beq
h_{\pm,s\Delta_j+\cdots \Delta_{2l}+k_1}= h_{\pm,k_1}+ s(\Delta_j +\cdots +\Delta_{2l})h'_{\pm,k_1} 
+ o(s^2) \ .
\label{scalingh}
\eeq
such that
\[
C_{s\Delta_1,...,s\Delta_{2l}}\sim C_0+s\sum_{j_1=1}^{2l}C_{j_1}\Delta_{j_1}+
\cdots +s^l\sum_{j_1,...,j_l=1}^{2l}
C_{j_1,...,j_l}\prod_{n=1}^l\Delta_{j_n}+o(s^{2l+1}) \ .
\]
Here it is important to notice that each $\Delta_j$ becomes a gradient term in real space, whereas a constant
term in $\Delta_j$ is diagonal in real space. Therefore, at least every second factor $\Delta_j$
(i.e., either $\Delta$'s with $j=1,3,...,2l-1$ or $j=2,4,...,2l$) must be
present, since otherwise multiple factors of $\psi_r$ or
${\bar\psi}_r$ at the same site $r$ would appear which gives zero due to the fact that these are 
Grassmann variables. 
Thus the leading behavior of the right-hand side of Eq. (\ref{scaling2}) under scaling is
\[
\sim s^{2l(2+\alpha)}s^{l-2}
\int C_{\Delta_1,...,\Delta_{2l}}\psi_{\Delta_1}{\bar\psi}_{\Delta_2}
\cdots \psi_{\Delta_{2l-1}}{\bar\psi}_{\Delta_{2l}}\delta(\Delta_1+\dots +\Delta_{2l})d^2\Delta_1\cdots d^2\Delta_{2l} \ .
\]
For $\alpha=-2$ this means that only terms with $l\le2$ are relevant for $s\sim0$.
Moreover, the $l=2$ term vanishes, since there are two contributions that cancel each other.
This can easily be seen in real-space representation:
\[
Tr(h_-\psi h_+{\bar\psi}h_-\psi h_+{\bar\psi}) =\sum_{r_1,...,r_4}
Tr_2(h_{-,r_1-r_2}h_{+,r_2-r_3}h_{-,r_3-r_4}h_{+,r_4-r_1})\psi_{r_2}{\bar\psi}_{r_3}\psi_{r_4}{\bar\psi}_{r_1} \ .
\]
The leading non-vanishing term is of order $s^2$. In this case, according to the gradient expansion,
every second term is diagonal and reads
\[
Tr_2(h_{-,0}h_{+,r_1-r_3}h_{-,0}h_{+,r_3-r_1})\psi_{r_1}{\bar\psi}_{r_3}\psi_{r_3}{\bar\psi}_{r_1}
+Tr_2(h_{-,r_1-r_2}h_{+,0}h_{-,r_2-r_1}h_{+,0})\psi_{r_2}{\bar\psi}_{r_2}\psi_{r_1}{\bar\psi}_{r_1} \ .
\]
After renaming the summation indices and exchanging of the Grassmann factors in the first term we get
\[
=\left[
-Tr_2(h_{-,0}h_{+,r_1-r_2}h_{-,0}h_{+,r_2-r_1})+Tr_2(h_{-,r_1-r_2}h_{+,0}h_{-,r_2-r_1}h_{+,0})\right]
\psi_{r_2}{\bar\psi}_{r_2}\psi_{r_1}{\bar\psi}_{r_1} \ .
\]
Now we use the fact that $h_{\pm}=\kappa_0\sigma_0\pm (\kappa_1\sigma_1+\kappa_2\sigma_2)$ in 
Eq. (\ref{hfourier}) and get from the sum of the two trace terms zero. This result implies that 
the loop expansion is asymptotically dominated by the term in Eq. (\ref{1loop}) (i.e. the loop with 
two corners) which give the propagator
\beq
\sum_r e^{-iq\cdot r}\langle\psi_r{\bar\psi}_0\rangle\sim 
\frac{1}{-2+C_q} \ .
\label{diffprop}
\eeq
Here $C_q$ can be expanded in powers of $q$ (cf. Appendix \ref{evpropagator}) as
\[
C_q=2-\frac{4\eta^2}{g\eta'}(\epsilon +Dq^2)
+ o(q^3)
\]
with the diffusion coefficient 
\beq
D:=-\frac{g\eta'}{2}\frac{\partial^2}{\partial q_k^2}\int
Tr_2\left[G_{0,22,k}(\epsilon+\eta)G_{0,22,k-q}(-\epsilon-\eta) \right]
d^2k |_{q=0} \ .
\label{defofd}
\eeq
Thus the propagator in Eq. (\ref{diffprop}) describes diffusion on large scales.
The $\epsilon$ term corresponds with the symmetry breaking parameter.
The latter does not need to be a scalar but can be any symmetry-breaking tensor in the Green's function, 
provided it allows us to write the two-particle Green's function in the form of Eq. (\ref{extendedgf}).

The matrix element of Eq. (\ref{me4}) reads with these expressions and the substitution
$\epsilon\to i\omega/2$
\[
\langle\langle\Phi_{\omega/2} |r_k^2|\Phi_{-\omega/2}\rangle\rangle_m
=-\frac{\partial^2}{\partial q_k^2}\frac{\eta'}{g}\frac{1}{i\omega/2+Dq^2}\Big|_{q=0}
=-8\frac{\eta'D}{g\omega^2} \ .
\]
We can also use the definition of $D$ in Eq. (\ref{defofd}), together with Eq. (\ref{cond2}),
to write
\beq
D=\frac{g\eta'}{2}\langle\Phi_{i\eta'}^0 |r_k^2|\Phi_{-i\eta'}^0\rangle
\label{diffcoeff1}
\eeq
and 
\beq
\langle\langle\Phi_{\omega/2} |r_k^2|\Phi_{-\omega/2}\rangle\rangle_m
=-\frac{\eta'^2}{(\omega/2)^2}\langle\Phi_{i\eta'}^0 |r_k^2|\Phi_{-i\eta'}^0\rangle \ ,
\label{gfrelation}
\eeq
where $|\Phi_{E}^0\rangle$ is the wave function of Eq. (\ref{evolution}), when the 
Hamiltonian is replaced by the translational-invariant Hamiltonian $H_0$. 
Moreover, the integration in Eq. (\ref{defofd}) gives for $\lambda\sim\infty$ 
(cf Appendix \ref{evmatrixelement}) 
\beq
D=\frac{ag\eta'}{(4\eta'^2+{\bar m}^2)\pi} \ \ 
(a=1\ \  {\rm for}\ {\rm MLG},\ \  a=2\ \  {\rm for}\ {\rm BLG})
\label{diffcoeff3}
\eeq
which implies
\beq
\langle\langle\Phi_{\omega/2} |r_k^2|\Phi_{-\omega/2}\rangle\rangle_m
\sim -\frac{8a\eta'^2}{\omega^2(4\eta'^2+{\bar m}^2)\pi}
 \ .
\label{matrixel5}
\eeq


\section{Discussion}
\label{discussion}

All our results are obtained for the charge neutrality point $E=0$, for mono- and for bilayer graphene. 
The main results are given in Eqs. (\ref{diffprop}), (\ref{gfrelation}), (\ref{diffcoeff1}), and (\ref{diffcoeff3}):
Eq. (\ref{diffprop}) connects the average two-particle Green's function with the two-particle Green's function of
the average Hamiltonian. A special consequence is 
Eq. (\ref{gfrelation}), which describes a relation between a disorder-averaged matrix element and the corresponding
matrix element of the pure system. Eq. (\ref{diffcoeff1}) connects the matrix element with the diffusion coefficient.
And finally, Eq. (\ref{diffcoeff3}) connects the diffusion coefficient with the one-particle scattering rate $\eta$.
\\

\no
{\it density of states:}
The average density of states is proportional to the diagonal element of the 
average one-particle Green's function $\langle (H+i\epsilon)^{-1}\rangle_m$. The latter
can be evaluated in saddle-point approximation from Eq. (\ref{SPE00}) as
\beq
\langle G(i\epsilon)\rangle_m 
\approx G_{0}(i\epsilon+i\eta) \ ,
\label{scba001}
\eeq
where the parameters $\eta$ (scattering rate) and $m_s$ are determined by the self-consistent
(or saddle-point) conditions of Eqs. (\ref{spea}), (\ref{speb}).
We then obtain 
$\rho_0\approx \eta/2\pi g$, where the scattering rate $\eta$ is a function of $g$ and ${\bar m}$, 
according to Eq. (\ref{scattrate}). The density of states has a semicircular form with respect to ${\bar m}$
\beq
\rho_0\approx \frac{\eta}{2\pi g}=\frac{1}{4\pi g}\sqrt{m_c^2-{\bar m}^2}\Theta(m_c^2-{\bar m}^2) \ ,
\label{dos01}
\eeq
where the radius of the semicircle $m_c$ is given in Eq. (\ref{gap11}).\\

\no
{\it diffusion:}
Scattering by the random gap term leads to diffusion, as explained in the loop expansion of Sect. \ref{loopexpansion}. 
The diffusion coefficient $D$ in Eq. (\ref{diffcoeff3}) depends only on $\eta'$.
This corresponds with the simple physical picture that diffusion decreases with an increasing scattering rate.
Diffusion breaks down when the symmetry is broken by the parameter $\epsilon$. This implies a maximal diffusion 
length $L_{\rm diff}=\sqrt{2D/\epsilon}$. The scale $L_{\rm eff}$ indicates that any symmetry-breaking term creates a finite 
diffusion length which limits diffusion to systems of linear size $L_{\rm eff}$. This length scale can be very large 
due to the large diffusion coefficient $D$ in MLG
\[
L_{\rm diff}\sim \frac{1}{2\sqrt{\pi}}\sqrt{\frac{g e^{\pi/g}}{\lambda\epsilon}} \ .
\]
In the case of BLG, however, it is much smaller because of the stronger scattering rate $\eta=m_c/2$
of Eq. (\ref{gap11}):
\[
L_{\rm diff}\sim \sqrt{\frac{2}{\pi\epsilon}} \ .
\]
\\

\no
{\it matrix element:}
The averaged matrix element $\langle\langle\Phi_{\omega/2} |r_k^2|\Phi_{-\omega/2}\rangle\rangle_m$
is an indicator of Anderson localization, since it diverges if the localization length is infinite.
According to Eq. (\ref{matrixel5}), the states $|\Phi_{\pm\omega/2}\rangle$ are delocalized at $\omega=0$.
On the other hand, the states are localized for $\omega\ne0$
with a decreasing localization length as one goes away from the NP. Such a behavior was also found for 
bond disorder in analytic \cite{ziegler08} and in numerical studies \cite{xiong07}.\\

\no
{\it relation between averaged and non-averaged Green's functions:}
In general, the average two-particle Green's function can be expressed by the function $C_q$ through Eq. (\ref{diffprop}). 
$C_q$ in Eq. (\ref{propagator2}) is a function of the Green's functions $G_0(\pm\eta')$, 
where the random Hamiltonian ${\hat H}$ is replaced by the average Hamiltonian $H_0$.
Since the average Hamiltonian is translational invariant, the function $C_q$ can be easily calculated.
This relation between the average two-particle Green's function and the self-consistent two-particle Green's function
\[
\sum_r e^{-iq\cdot r}Tr_2\left[
\langle  G_{r0}(-i\epsilon)G_{0r}(i\epsilon)\rangle_m
\right]
\approx \frac{1}{-2+C_q}
\]
can be considered as a generalization of the self-consistent Born approximation of the one-particle Green's
function in Eq. (\ref{scba001}). 
Like in the latter case,
the averaging process leads to a change of energies $\epsilon\to\eta'$ (i.e. a replacement
of the symmetry-breaking parameter by the scattering rate). 
A consequence for the matrix element is Eq. (\ref{gfrelation}), which means a simple scaling relation 
between the average matrix element and the matrix element of the average translational-invariant Hamiltonian 
$H_0$.
(The scale $\eta'$, however, is not free but fixed by the disorder average through the saddle-point 
equation (\ref{SPE00}).)
This provides an interesting and useful relation between averaged and non-averaged Green's functions.
Moreover, in the relation of the matrix elements
there is an extra prefactor $-\eta'^2/(\omega/2)^2$. This is important for the transport properties, 
since it provides the delocalization of states at $\omega=0$ and
it cancels the factor $\omega^2$ in the conductivity of Eq. (\ref{cond0b}). The relation in
Eq. (\ref{gfrelation}) can also be understood as
a factorization of the averaged matrix element into a product of a power law (i.e. $\sim\omega^{-2}$)
and a smooth scaling function $\eta'^2\langle\Phi_{i\eta'}^0 |r_k^2|\Phi_{-i\eta'}^0\rangle$.\\

\no
{\it conductivity:}
The conductivity of Eq. (\ref{cond0b}) is calculated from the matrix element in Eq. (\ref{matrixel5})
and gives 
\beq
\sigma_0(\omega)\sim \frac{4a\eta'^2}{\pi(4\eta'^2+{\bar m}^2)}\Theta(m_c^2-{\bar m}^2)\frac{e^2}{h} \ .
\label{condfinal}
\eeq
It is remarkable that $\eta'$ drops out for ${\bar m}=0$ which gives a frequency-independent result
\[
\sigma_0(\omega)\sim \frac{a}{\pi}\frac{e^2}{h} \ .
\]
A frequency-independent conductivity was also found for a random vector potential \cite{ziegler08}.
In the absence of disorder a constant $\sigma(\omega)$ was found, with a different value though 
\cite{ziegler07,stauber08}. 
The difference is due the fact that the expression in Eq. (\ref{cond0b}) is only a contribution due to 
interband scattering from the total Kubo formula (for details cf Ref. \cite{ziegler08}).

{\it DC conductivity:}
For $\omega\sim0$ the parameter $\eta'$ is replaced by the scattering rate $\eta$ of Eq. (\ref{gap11}). 
The resulting DC conductivity reads
\beq
\sigma_0(\omega\sim0)\sim \frac{4a\eta^2}{\pi(4\eta^2+{\bar m}^2)}\frac{e^2}{h}
=\frac{a}{\pi}\left(1-\frac{{\bar m}^2}{m_c^2}\right)\Theta(m_c^2-{\bar m}^2)\frac{e^2}{h} \ .
\label{cond5}
\eeq

Our knowledge of the diffusion coefficient $D$ in Eq. (\ref{diffcoeff3}) and the density of states $\rho_0$ 
in Eq. (\ref{dos01}) 
allows us to evaluate the DC conductivity alternatively through the Einstein relation:
\[
\sigma(\omega\sim0)\propto \rho D\frac{e^2}{h}
\approx \frac{a}{8\pi^2}\left(1-\frac{{\bar m}^2}{m_c^2}\right)\Theta(m_c^2-{\bar m}^2)\frac{e^2}{h}  \ .
\]
This agrees with Eq. (\ref{cond5}), except for a constant factor.
\begin{figure}
\begin{center}
\includegraphics[width=7.5cm,height=6.5cm]{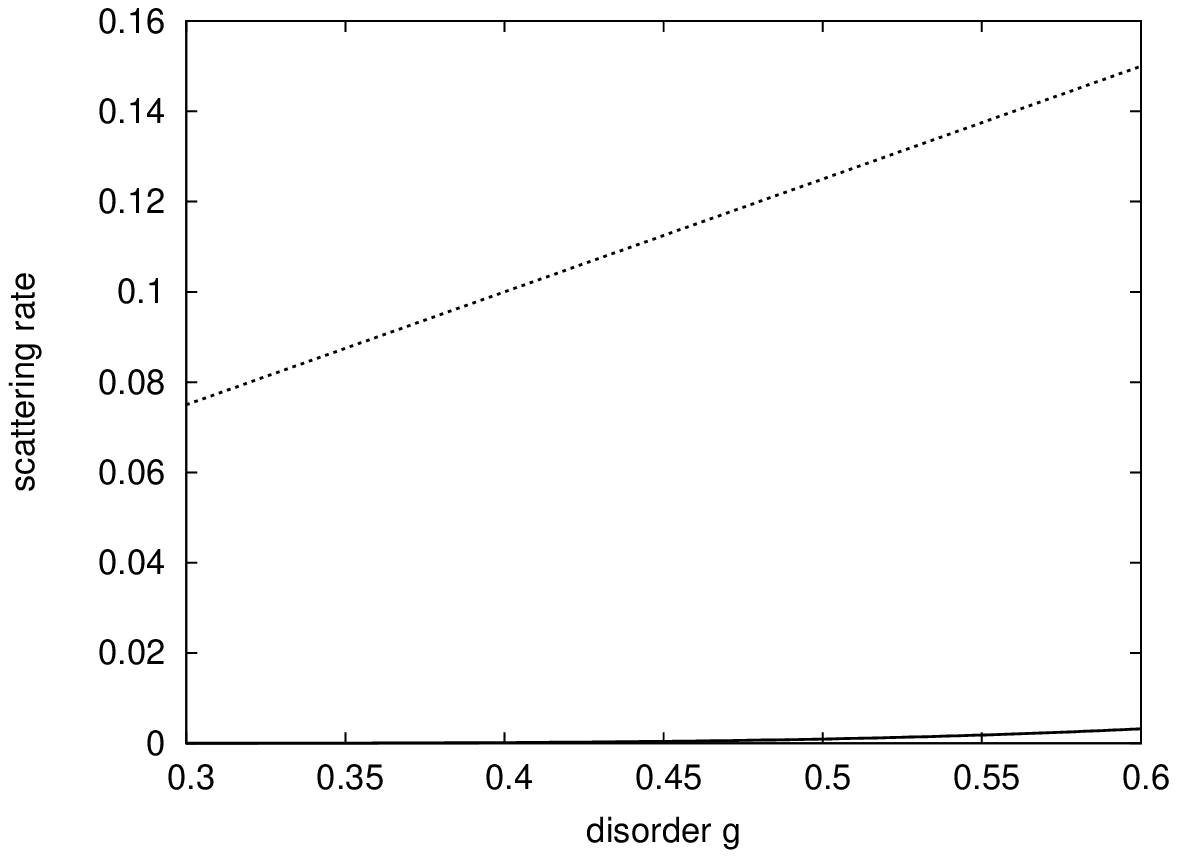}
\includegraphics[width=7.5cm,height=6.5cm]{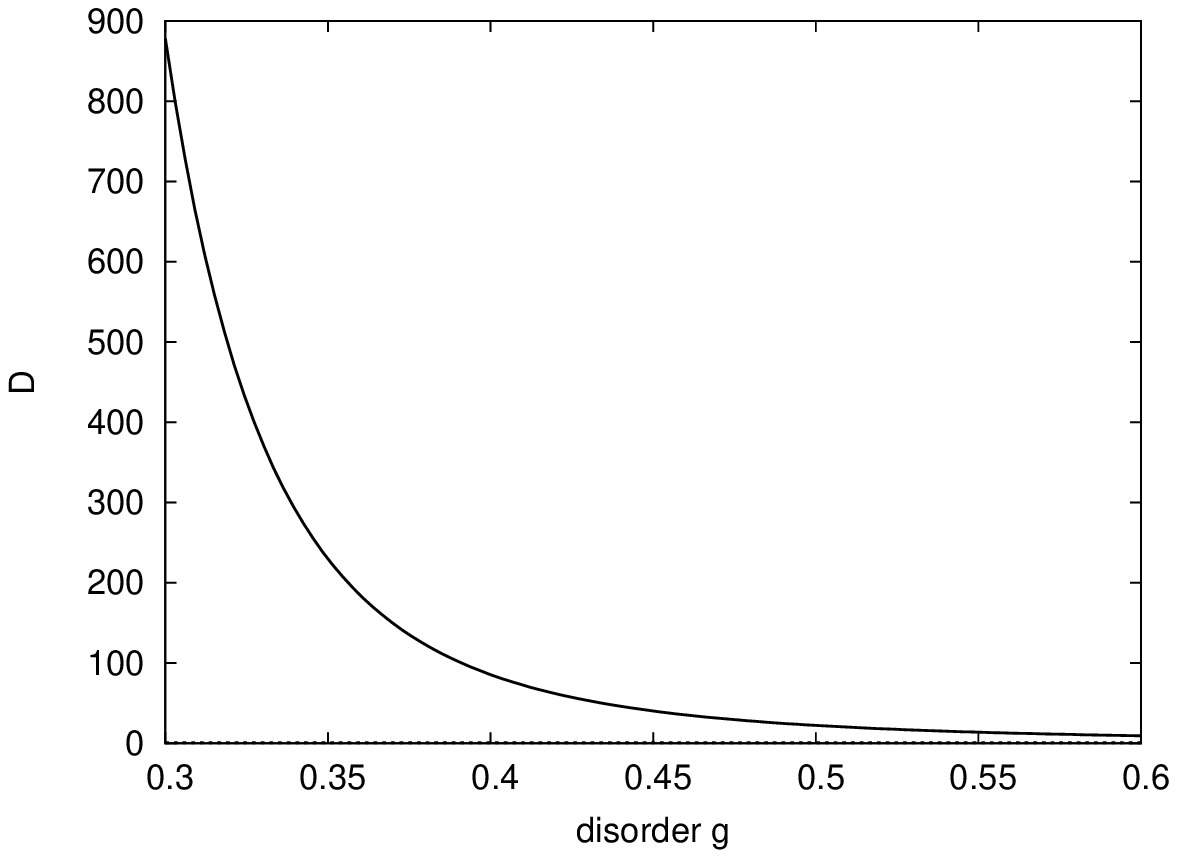}
\caption{
Scattering rate $\eta$ and diffusion coefficient $D$ for ${\bar m}=0$ in the case of monolayer graphene 
(full curves) and bilayer graphene (dashed curves) versus the variance $g$ of the random symmetry-breaking potential. 
The diffussion coefficient of bilayer graphene is so small ($D\sim 2/\pi$) such that it cannot be 
distinguished from the $g$ axis.}
\label{plots}
\end{center}
\end{figure}

It is important to notice that the conductivity at ${\bar m}=0$ does not depend on the variance $g$ of the random SBP.
This indicates that this quantity is robust against random fluctuations in graphene. In particular,
we could have started from the action in Eq. (\ref{effaction}) and treated the interaction term in perturbation
theory in powers of $g$ to obtain the same result. This idea was indeed employed in Ref. \cite{dcconductivity} and gave 
the same value for the minimal conductivity. However, it is not possible to obtain a non-zero critical value $m_c$
in the case of MLG, since all orders of the expansion of $m_c$ in Eq. (\ref{gap11}) give zero. This is one of the
reasons why we have not used the perturbation theory in $g$ here but the loop expansion of Sect. \ref{loopexpansion}.

\section{Conclusion}
\label{conclusions}

The physics of the random gap model is characterized by a discrete symmetry of the Hamiltonian
and a continuous symmetry of the two-particle Green's function. For the disorder-averaged two-particle
Green's function the continuous symmetry is represented by a fermionic degree of freedom. Since the
symmetry is spontaneously broken, the resulting massless fermion mode controls the properties
on large scales. An effective action is derived for the massless fermion mode and a loop expansion is
employed to extract the dominant large-scale contribution. It is found that the shortest loops
are in control of the large scales, leading to diffusion. An explicitly broken symmetry generates
a finite diffusion length $L_{\rm diff}$ such that diffusion is possible only on length scales less
than $L_{\rm diff}$. 

Although our models of mono- and bilayer graphene share the same type of symmetries and symmetry
breaking, the quantitative properties are quite different, since scattering is much stronger in
bilayer graphene (cf Fig. \ref{plots}). 
For instance, the diffusion coefficient $D$ is very large for monolayer graphene, namely
\[
D\propto ge^{\pi/g}
\]
for average symmetry-breaking potential ${\bar m}=0$,
because the low density of states at the neutrality point does not provide much scattering.
This means that transport in monolayer graphene is practically ballistic if disorder is not too strong.
In the case of bilayer graphene, however, scattering is much stronger because of a large density of 
states at the neutrality point, leading to a constant diffusion coefficient $D\sim 2/\pi$ for ${\bar m}=0$. 
This also implies a large diffusion length scale $L_{\rm diff}$ for monolayer graphene since
$L_{\rm diff}\propto \sqrt{D}$. 

All physical quantities of our discussion (i.e. the average density of states, the diffusion coefficient, and
the matrix element of the position operator) depend on the model parameters only through the 
one-particle scattering rate $\eta$. An exceptional case is the conductivity for vanishing average 
symmerty-breaking potential which is independent of the model parameters at all and has the value $e^2/\pi h$ for
monolayer graphene and $2e^2/\pi h$ for bilayer graphene (up to a factor 4 for spin and valley degeneracy). 
This implies
a frequency-independent microwave conductivity. On the other hand, an increasing average symmetry-breaking potential 
${\bar m}$ reduces continuously the conductivity as well as the diffusion coefficient.
The continuous behavior of the conductivity with respect to gap opening is similar to a recent 
experimental observation by Adam et al. \cite{adam08}.

\begin{acknowledgments}
This project was supported by a grant from the Deutsche Forschungsgemeinschaft.
\end{acknowledgments}

\appendix

\section{Integration over the nonlinear field}
\label{invmeas}

We consider the matrix expansion at fixed site $r$:
\[
{\hat Q}=Q_{11}+Q_{12}+\cdots
\]
where $\{ Q_{ij} \}$ is a basis for the matrix ${\hat Q}$.
In the integral
\[
I_1=\int f(Q_{11,11},Q_{12,12},Q_{21,12},...)
\]
$Q_{11,11}$ is replaced by the nonlinear term $Q_{11,11}+1+Q_{12,12}Q_{21,12}$ which is
created by the diagonal matrix elements of ${\hat U}^2$ from the saddle-point manifold.
This leads to the new integral
\[
I_2=\int f(Q_{11,11}+1+Q_{12,12}Q_{21,12},Q_{12,12},Q_{21,12},...) \ .
\]
An expansion in terms of the Grassmann variable $Q_{12,12}Q_{21,12}$ gives
\[
I_2=\int f(Q_{11,11}+1,Q_{12,12},Q_{21,12},...)
+\int Q_{12,12}Q_{21,12}f'(Q_{11,11}+1,Q_{12,12},Q_{21,12},...)
\]
The second term vanishes at the $Q_{11,11}$ integration boundaries.
Moreover, the shift in the first term by 1 can be removed, since the integration of $Q_{11,11}$
goes from $-\infty$ to $\infty$. This gives
\[
I_2=\int f(Q_{11,11},Q_{12,12},Q_{21,12},...)=I_1 \ .
\]

\section{Diffusion propagator}
\label{evpropagator}

$C_q$ is defined in Eq. (\ref{coeffloop}):
\[
C_q=\int Tr_2(h_{-,k}h_{+,k-q})d^2k
=\int Tr_2
[\sigma_0-2i\eta G_{0,22,k}(i\epsilon+i\eta)][\sigma_0+2i\eta G_{0,22,k-q}(-i\epsilon-i\eta)]d^2k
\]
\[
=\int\left\{
2+2i\eta Tr_2[G_{0,22,k-q}(-i\epsilon-i\eta)-G_{0,22,k}(i\epsilon+i\eta)]
+4\eta^2 Tr_2\left[G_{0,22,k}(i\epsilon+i\eta)G_{0,22,k-q}(-i\epsilon-i\eta) \right]
\right\}d^2k
\]
\beq
=2+\int\left\{
2i\eta Tr_2[G_{0,22,k-q}(-i\epsilon-i\eta)-G_{0,22,k}(i\epsilon+i\eta)]
+4\eta^2 Tr_2\left[G_{0,22,k}(i\epsilon+i\eta)G_{0,22,k-q}(-i\epsilon-i\eta) \right]
\right\}d^2k \ ,
\label{propagator}
\eeq
since the $k$ integral is normalized. The Green's function reads
\[
G_{0,22,k}(i\epsilon+i\eta)=-\frac{1}{(\epsilon+\eta)^2+h_1^2+h_2^2}
[i(\epsilon+\eta)-h_1\sigma_1+h_2\sigma_2] \ .
\]
Using the saddle-point equation (\ref{SPE00}) with $\eta'=\eta+\epsilon$, we have
\[
\eta=\pm igTr_2[G_{0,22,rr}(\pm i\eta')] \ .
\]
This implies
\[
Tr_2[G_{0,22,rr}(-i\eta')-G_{0,22,rr}(i\eta')]=2i\eta/g \ ,
\]
such that
\beq
C_q=2-\frac{4\eta^2}{g}+4\eta^2 \int
Tr_2\left[G_{0,22,k}(i\eta')G_{0,22,k-q}(-i\eta') \right]d^2k \ .
\label{propagator2}
\eeq
The second term can be expanded in powers of $q$:
\[
C_q=2-\frac{4\eta^2}{g}+4\eta^2 \int
Tr_2\left[G_{0,22,k}(i\eta')G_{0,22,k}(-i\eta') \right]d^2k
\]
\beq
+2\eta^2 q_k^2\frac{\partial^2}{\partial q_k^2}\int
Tr_2\left[G_{0,22,k}(i\eta')G_{0,22,k-q}(-i\eta') \right]
d^2k |_{q=0}+o(q^3) \ .
\label{secondterm}
\eeq
Since $G_0$ satisfies the following relations
\[
G_{0}(i\eta')G_{0}(-i\eta')
=(i\eta'+h_0)^{-1}(-i\eta'+h_0)^{-1}=(\eta'^2+h_0^2)^{-1}
\]
and
\[
G_{0}(i\eta')-G_{0}(-i\eta')
=(i\eta'+h_0)^{-1}-(-i\eta'+h_0)^{-1}
=-2i\eta'(\eta'^2+h_0^2)^{-1} \ ,
\]
we obtain
\[
Tr_2 G_{0}(i\eta')G_{0}(-i\eta')
=\frac{i}{2{\eta'}}Tr_2[G_{0}(i\eta')-G_{0}(-i\eta')] \ .
\]
This allows us to write for the third term in Eq. (\ref{secondterm})
\[
4\eta^2 \int
Tr_2\left[G_{0,22,k}(i\eta')G_{0,22,k}(-i\eta') \right]d^2k
\]
\[
=4\eta^2 Tr_2[G_{0,22}(i\eta')G_{0,22}(-i\eta')]_{rr}
=2i\frac{\eta^2}{\eta'}Tr_2[G_{0,22}(i\eta')-G_{0,22}(-i\eta')]_{rr}
=4\frac{\eta^3}{g\eta'} \ .
\]
This gives
\[
C_q=2-\frac{4\eta^2}{g\eta'}\left[\epsilon
-q_k^2\frac{g\eta'}{2}\frac{\partial^2}{\partial q_k^2}\int
Tr_2\left[G_{0,22,k}(i\eta')G_{0,22,k-q}(-i\eta') \right]
d^2k |_{q=0}\right] +o(q^3) \ .
\]
Then the prefactor $D$ of the $q_k^2$ term reads
\[
D:=-\frac{g\eta'}{2}\frac{\partial^2}{\partial q_k^2}\int
Tr_2\left[G_{0,22,k}(i\eta')G_{0,22,k-q}(-i\eta')\right]
d^2k |_{q=0} \ .
\]

\section{Evaluation of the matrix element of $|\Phi_{\pm i\eta'}^0\rangle$}
\label{evmatrixelement}

The matrix element with respect to the average Hamiltonian $\langle H\rangle_m$ of MLG gives
\[
\langle\Phi_{i\eta'}^0 |r_k^2|\Phi_{-i\eta'}^0\rangle
=4(\eta'^2+{\bar m}^2/4)\int_0^\lambda\frac{k}{(\eta'^2+{\bar m}^2/4+k^2)^3}\frac{dk}{2\pi}
\sim \frac{1}{2\pi(\eta'^2+{\bar m}^2/4)}
\]
for $\lambda\sim\infty$, and of BLG
\[
\langle\Phi_{i\eta'}^0 |r_k^2|\Phi_{-i\eta'}^0\rangle
=16(\eta'^2+{\bar m}^2/4)\int_0^\lambda\frac{k^3}{(\eta'^2+{\bar m}^2/4+k^4)^3}\frac{dk}{2\pi}
\sim \frac{1}{\pi(\eta'^2+{\bar m}^2/4)} \ .
\]


\begin{thebibliography}{99}

\bibitem{novoselov05}
K.S. Novoselov, A.K. Geim, S.V. Morozov, D. Jiang, M.I. Katsnelson, I.V. Grigorieva, S.V. Dubonos,
A.A. Firsov, Nature {\bf 438}, 197 (2005)

\bibitem{zhang05}
Y. Zhang, Y.-W. Tan, H.L. Stormer, P. Kim, Nature {\bf 438}, 201 (2005)

\bibitem{geim07}
A.K. Geim and K.S. Novoselov, Nature Materials, {\bf 6}, 183 (2007)

\bibitem{tan07}
Y.-W. Tan, Y. Zhang, K. Bolotin, Y. Zhao, S. Adam, E.H. Hwang, S. Das Sarma, H.L. Stormer, P. Kim,
Phys. Rev. Lett. {\bf 99}, 246803 (2007) 

\bibitem{chen08}
J.H. Chen, C. Jang, M.S. Fuhrer, E.D. Williams, M. Ishigami,
Nature Physics {\bf 4}, 377 (2008)

\bibitem{morozov08}
S.V. Morozov, K.S. Novoselov, M.I. Katsnelson, F. Schedin, D.C. Elias, J.A. Jaszczak, A.K. Geim,
Phys. Rev. Lett. {\bf 100}, 016602 (2008)

\bibitem{elias08}
D.C. Elias, R.R. Nair, T.M.G. Mohiuddin, S.V.Morozov, P. Blake, M.P.H alsall,  
A.C. Ferrari, D.W. Boukhvalov, M.I. Katsnelson, A.K. Geim, K.S. and Novoselov,
Science  {\bf 323}, 610 (2009)  

\bibitem{ohta06}
O. Taisuke, A. Bostwick, T. Seyller, K. Horn, E. Rotenberg,
Science {\bf 18}, Vol. 313, 951

\bibitem{gorbachev08}
R.V. Gorbachev, F.V. Tikhonenkoa, A.S. Mayorova, D.W. Horsella and A.K. Savchenkoa,
Physica E {\bf 40}, 1360 (2008)

\bibitem{oostinga08}
J.B. Oostinga, H.B. Heersche, X. Liu, A.F. Morpurgo, L.M.K. Vandersypen,
Nature Materials {\bf 7}, 151 (2008)

\bibitem{mccann06b}
E. McCann and V.I. Fal'ko, Phys. Rev. Lett. {\bf 96}, 086805 (2006); 
E. McCann, Phys. Rev. B {\bf 74}, 161403(R) (2006)

\bibitem{castro08}
E.V. Castro, N.M.R. Peres, J.M.B. Lopes dos Santos, F. Guinea, and A.H. Castro Neto,
J. Phys.: Conf. Ser. 129 012002 (2008)

\bibitem{cheianov07}
V.V. Cheianov, V.I. Fal'ko, B.L. Altshuler, and I.L. Aleiner,
Phys. Rev. Lett {\bf 99}, 176801 (2007)

\bibitem{zhang08}
Y.-Y. Zhang, Jiangping Hu, B.A. Bernevig, X.R. Wang, X.C Xie, W.M Liu,
Phys. Rev. Lett. {\bf 102}, 106401 (2009)

\bibitem{xiong07}
S.-J. Xiong and Y. Xiong, Phys. Rev. B {\bf 76}, 214204 (2007)

\bibitem{ziegler08}
K. Ziegler, Phys. Rev. B {\bf 78}, 125401 (2008) 

\bibitem{dcconductivity}
A.W.W. Ludwig, M.P.A. Fisher, R. Shankar, G. Grinstein, Phys. Rev. B {\bf 50}, 7526 (1994);
E. Fradkin, Phys. Rev. B {\bf 33}, 3263 (1986) 


\bibitem{ziegler97}
K. Ziegler, Phys. Rev. B {\bf 55}, 10661 (1997); Phys. Rev. Lett. {\bf 80}, 3113 (1998)

\bibitem{ziegler09a}
K. Ziegler, Phys. Rev. Lett. {\bf 102}, 126802 (2009) 

\bibitem{suzuura02}
H. Suzuura and T. Ando, Phys. Rev. Lett. {\bf 89}, 266603 (2002)

\bibitem{peres06}
N.M.R. Peres, F. Guinea, and A.H. Castro Neto, Phys. Rev. 
B {\bf 73}, 125411 (2006)

\bibitem{khveshchenko06}
D. Khveshchenko, Phys. Rev. Lett. {\bf 97}, 036802 (2006)

\bibitem{mccann06}
E. McCann et al., Phys. Rev. Lett. {\bf 97}, 146805 (2006)

\bibitem{yan08}
X.-Z. Yan and C.S. Ting, Phys. Rev. Lett. {\bf 101}, 126801 (2008)

\bibitem{koshino06}
M. Koshino and T. Ando, Phys. Rev. B {\bf 73}, 245403 (2006)

\bibitem{morozov06}
S.V. Morozov et al., 
Phys. Rev. Lett. {\bf 97}, 016801 (2006)

\bibitem{castroneto07b}
A.H. Castro Neto, F. Guinea, N.M.R. Peres, K.S. Novoselov, and A.K. Geim, 
Rev. Mod. Phys. 

\bibitem{meyer07}
J.C. Meyer, A. K. Geim, M. I. Katsnelson, K. S. Novoselov, T. J. Booth, S. Roth, 
Nature {\bf 446}, 60 (2007) 

\bibitem{anderson58}
P.W. Anderson, Phys. Rev. {\bf 109}, 1492 (1958)

\bibitem{zitter}
J. Cserti and G. D\'avid, 
Phys. Rev. B {\bf 74}, 172305 (2006);
M.I. Katsnelson, 
Eur. Phys. J. B {\bf 51}, 157-160 (2006);
M.I. Katsnelson and K. S. Novoselov, Solid State Commun. {\bf 143}, 3 (2007); 
T.M. Rusin and W. Zawadzki, 
Phys. Rev. B {\bf 76}, 195439 (2007);
U. Z\"ulicke, J. Bolte, and R. Winkler, New J. Phys. {\bf 9}, 355 (2007);
J. Schliemann, New J. Phys. {\bf 10}, 043024 (2008) 

\bibitem{negele}
J.W. Negele and H. Orland, {\sl Quantum Many-Particle Physics}, Addison-Wesley, New York (1988) 

\bibitem{ziegler07}
K. Ziegler, Phys. Rev. B {\bf 75}, 233407 (2007)

\bibitem{stauber08}
T. Stauber, N.M.R. Peres, and A.K. Geim, 
Phys. Rev. B 78, 085432 (2008)

\bibitem{adam08}
S. Adam, S. Cho, M.S. Fuhrer, and S. Das Sarma, Phys. Rev. Lett. {\bf 101}, 046404 (2008) 

\end{thebibliography}
\end{document}